\documentclass[11pt]{article}

\usepackage{geometry}
\usepackage{fullpage}
\usepackage{graphicx}
\usepackage{pdflscape}
\usepackage{multirow,multicol}
\usepackage{ragged2e}

\usepackage{dcolumn}
\usepackage{booktabs,calc}
\usepackage{longtable}
\usepackage{array}
\usepackage{paralist}
\usepackage{verbatim}
\usepackage{subfig}
\usepackage{setspace}
\usepackage{amsmath}
\usepackage{amssymb}
\usepackage{natbib}
\usepackage[linktocpage=true, colorlinks=true, linkcolor=blue, citecolor=black]{hyperref}
\usepackage{enumerate}
\usepackage{authblk}
\usepackage{placeins}

\begin{document}
\title{Differentiation in a Two-Dimensional Market with Endogenous Sequential Entry\thanks{Corresponding author email: jmichler@arizona.edu. Preliminary draft, do not cite without permission. We greatly appreciate the assistance of and helpful comments from conference participants at the AAEA Annual Meetings in Seattle, WA. We are solely responsible for any errors or misunderstandings.}}

\author[a]{Jeffrey D. Michler}
\author[b]{Benjamin M. Gramig}
\affil[a]{\small \emph{Department of Agricultural and Resource Economics, University of Arizona, Tucson, USA}}
\affil[b]{\small \emph{Dept. of Agricultural and Consumer Economics, University of Illinois, Urbana, USA}}

\date{March 2021}
\maketitle
	
	\begin{center}\begin{abstract}
			\noindent Previous research on two-dimensional extensions of Hotelling's location game has argued that spatial competition leads to maximum differentiation in one dimensions and minimum differentiation in the other dimension. We expand on existing models to allow for endogenous entry into the market. We find that competition may lead to the min/max finding of previous work but also may lead to maximum differentiation in both dimensions. The critical issue in determining the degree of differentiation is if existing firms are seeking to deter entry of a new firm or to maximizing profits within an existing, stable market.
		\end{abstract}\end{center}

		{\small \noindent\emph{JEL Classification}: D43, L11, L13
			}
		
\newpage
\doublespacing
\section{Introduction}\label{sec:introduction}
Despite adages on the importance of location for business profitability, little work has been done to expand rudimentary one-dimensional economic models of location choice into a more realistic two-dimensions. Location choice in these models is often interpreted as taking place in product space instead of physical space. Yet, despite the multi-dimensional characteristics of most products, existing models still focus on product differentiation along a single dimension. The purpose of this paper is to take an exploratory step in expanding models of product differentiation or location choice into a second dimension.

The location game was first developed by Hotelling (1929). The game is played between $n$ firms in one-dimensional space with consumers uniformly distributed along an interval normalized to $[0,1]$. Firms are homogeneous with zero production costs. Consumers purchase a single unit of the good from the firm closest to their location at a uniform price, $p$. If a consumer is equidistant between two firms, they split their purchases equally between the two firms. Firms enter the market simultaneously and attempt to maximize profit. There is no single solution to Hotelling’s game. Rather, solutions are dependent on the number of firms in the game. It is worth noting that solution to Hotelling’s game, while optimal for the profit maximizing firm, are not socially optimal.

A socially optimal configuration can be achieved by imposing the restriction that firms, once located, cannot move. Additionally, entry in this game is generally assumed to be sequential. Independently, Hay (1976) and Rothschild (1976) derived solutions for the sequential entry, fixed location game. When location is fixed and entry sequential, firms will space themselves at uniform distances along the line. This solution is distinct from the Hotelling solution of agglomeration in that it is not only stable but socially optimal. The mobility constraint is enough to turn an inefficient equilibrium into an efficient one. For the most part, the literature has neglected to explore the implications of this transition in equilibriums.

A few papers have attempted to expand the Hotelling model into two dimensions but this is a small subset of the papers devoted to spatial competition (see Eiselt et al. 1993 for a good, though dated, review of this literature). Since Hakimi (1983), most location games in real space have defined the problem in terms of medianoid and centroid - polygon shaped areas to be maximized. These problems are computationally complex and without some fixed starting point or limiting assumption they tend to fall into the NP classification of problems. The limiting assumption that location, once chosen, is fixed greatly reduce the problem's degree of difficulty. Lane (1980) and Veendorp and Majeed (1995) each have taken this fixed location approach in order to simplify the multi-dimensional problem. Yet, in both cases while the model accommodates two-dimensions, the authors make additional assumptions to reduce the problem to the classic single dimension.

In both one-dimensional and two-dimensional space, when firms enter simultaneously and are mobile, the location problem is an indefinitely repeated game and generally lacks pure strategies. When firms enter sequentially and are mobile, no pure strategy exists on the line and solutions are generally unattainable in two-dimensional space. However, when firms enter sequentially and are fixed, a pure strategy exists on the line. Furthermore, that strategy results in a stable and sometime socially optimal equilibrium. In the model that follows, we adopt the fixed firm approach to simplify and solve for a stable equilibrium to the location game in two-dimensions. Furthermore, we examine the equilibrium to determine if it is socially optimal.

Our results resemble those derived by Neven (1987) and Gotz (2005) for an endogenous sequential entry location game. As opposed to the Hotelling Model, in two-dimensions firms do not agglomerate but attempt to maximize their differentiation from other firms. As opposed to Veendorp and Majeed (1995), who argue that firms would maximize differentiation in one dimension while minimizing differentiation in a second dimension, we find that firms will sometimes pursue maximimum differentiation in both dimensions. The characterization of the firm's location (max or min differentiation) depends on the potential for subsequent entrants. Similar to the results for a single dimension in Neven (1987) and Gotz (2005), we find that firms use location as a deterrence mechanism. In two-dimensions, firms maximize their distance from other firms along all dimensions when there is no threat of subsequent entry while "filling in" gaps in the market space when the threat of a subsequent entrant arises.

In section 2, we lay out the model for location choice in two-dimensions. In section 3, we present the results from our numerical simulation, showing the location choice is a function of the market size . Section 4 contains our discussion of the results while section 5 concludes.

\section{The Model}\label{sec:model}
We begin by defining the space over which the location choice is made as well as how firms are organized in that space. Consider a finite number of homogeneous firms. Identify each firms as $i \in I=(1,2,...,n)$. The location of $i$ is $(x_{i1},x_{i2})=\bf{x_i}$ in subspace $S$ of two-dimensional Cartesian space. Let $S$ be a square of unit length. Let $\bf{X}=(\bf{x_1^T},\bf{x_2^T},..., \bf{x_n^T})^T$ be a vector of $2n$ elements where \textbf{T} is the transpose operator. We define a configuration as a combination of firm locations in the subspace $C=(\textbf{X} , S)$. 

\subsection{Consumers}\label{sec:consumers}
For consumers, we modify the model developed by Veendorp and Majeed (1995).  In their paper, the authors set up the problem from the firm's perspective, where the firm maximizes the difference between price and the cost of delivering the good from its location at $\bf{x_i}$ to the consumer located at $\bf{\hat x}$. In order to facilitate comparability with the recent literature, especially those of Economides, et al. (2002) and Gotz (2005), we redefine the problem from the consumer's perspective.

There is a continuum of consumers uniformily distributed over subspace $S$ with density $M$. Each consumer expresses unit demand so that the density, $M$, is a measure of the market size. Let consumer location in subspace $S$ be defined as $(\hat x_1,\hat x_2)=\bf{\hat x}$. One can interpret $\bf{\hat x}$ as either the physical location of a type of consumers or as the most preferred product variety from a set of product varieties $\bf{X}$.

While consumers are heterogeneous by type, they share the same utility function. Consumer's utility from purchasing a good from firm $i$ is:

$$U_{\mathbf{\hat x}}^i = a - t \sqrt{(\mathbf{x}_i - \mathbf{\hat x})^2} - p_i = a - t \sqrt{(x_{i1} - \hat x_1)^2 + (x_{i2} - \hat x_2)^2}- p_i  \eqno (1)$$

\noindent where $a$ is the reservation value for the good, which is common to all consumers and sufficiently large so as to assure all consumers purchase the good. $p_i$ is the price chareged for the good charged by firm $i$. Transportation costs, $t$, are normalized to 1, without loss of generality.

\subsection{Firms}\label{sec:firms}
To solve the location problem we must first define the boundaries of the market to which a firm sells. The boundary points allow us to determining the market size for each firm, and thus the quantity of the good a firm sells. The boundary points are defined by the consumer who is indifferent between purchasing the good from firm $i$ or firm $j$. Following Neven's (1987) one-dimensional approach, define $\alpha_{i,j} = (\mathbf{x}_i , \mathbf{x}_j , p_i, p_j)$ as the consumer indifferent between $\mathbf{x}_i$ at price $p_i$ and $\mathbf{x}_j$ at price $p_j$, such that $U_{\alpha_{i,j}}^i = U_{\alpha_{i,j}}^j$. The model's set up allows for differentiation along both axes and contains as a special case the one dimensional model of Nevens (1987) and Gotz (2005). 

In the single dimensional model, a firm's market share can be completely characterized by just two consumers. If Firms $i,j,k$ are located sequentially along the line segment [0,1], then firm $i$'s market share is $D_i = M(\alpha_{i,j} -0)$ while firm $j$'s market share is  $D_j = M(\alpha_{j,k} - \alpha_{i,j})$  and firm $k$'s market share is  $D_k = M(1- \alpha_{j,k})$. In two dimensions, a firm's market share is defined by a vector of indifferent consumers that marks the boundary between two firms. Furthermore, while along the line a firm will have no more than two neighboring competitors, on the plane a firm may have $n-1$ neighboring competitors.

Because of these complications, the simplest method for defining a firm's market space in two dimensions is with a Voronoi Polygon. For firm $i$, every point $\bf{x}$ that is closer to $\mathbf{x}_i$ than to $\mathbf{x}_j$ is in the Voronoi polygon $V(\mathbf{x}_i)$.

$$V( \mathbf{x}_i)=\{ \mathbf{x} : \ \parallel \mathbf{x}_i - \mathbf{x} \parallel \ \leq \ \parallel \mathbf{x}_j - \mathbf{x} \parallel , \rm \  j \in I \} \eqno (3)$$

\noindent Given our assumptions regarding the distribution of consumers, firm $i$'s market share is

$$V_i= \alpha M \ \ \ \text{where} \ \ \ \alpha = \{ \alpha : \alpha = \alpha_{i,j} , j \in I \} \eqno (4)$$

\noindent What Equation (4) says is that one can measure the size of firm $i$'s market by taking the convex hull of the market space, as defined by the vectors of consumers indifferent between purchasing the good from $i$ or all of $i$'s $j$ competitors, and multiplying that by the density of consumers.

The firm's problem then is the following:

$$\max_{p_i , \mathbf{x}_i} \pi_i = (p_i - c)V_i - F \eqno (5)$$

\noindent where $c$ and $F$ are the firm's marginal and fixed costs of production. We can set $c=0$ without effecting the generalizability of our results. Each firm chooses a price and a location to maximize its profit but since price is a function of the prices and location of other firms, as well as its own location, the profit function can be re-written in terms of the single choice variable, $\mathbf{x}_i$.

\subsection{The Location Choice Problem}\label{sec:location}

The location game proceeds as follows: In the first stage, each firm enters sequentially and chooses a location. After all firms have entered, the second stage begins in which firms compete on price. While sequential entry is a natural candidate for a dynamic model, we adopt Neven's (1987) assumption that firms' discount rates are low so that profits earned during the transitory entry periods are negligible when compared to long run profits. While this assumption limits the possibly interesting dynamic aspect of the model, it lends justification for the fixed location assumption. Profits to be gained in the short run by a move from one location to another are outweighed by the fixed cost.

Consider, then, the problem for firm $N$, the last firm to enter. The firm can see the locations of all $N-1$ firms. Since prices are a function of location, firm $N$ can determine the prices that will prevail at the selected locations. This reduces firm $N$'s problem to a single unknown and gives rise to a solution, $\mathbf{x}_N (\mathbf{x}_1, . . . , \mathbf{x}_{N-1})$, that specifies the location of firm $N$ in terms of the locations of all other firms. Thus, each firm has only to solve a single equation in a single unknown. Using backward induction we continue this process from the last firm to enter to the first firm to enter. In each case, the location of preceding firms is known and the location of subsequent firms will be given by the choice function that we have already solved.

Lane (1980) showed that in one dimension, for a given set of locations a unique price equilibrium exists. However, there is no proof for the existence of a unique location equilibrium. This is because the location choice function is not necessarily a continuous function. The lack of a proof for a unique location equilibrium does not mean location equilibriums do not exist, only that we cannot demonstrate that a given equilibrium is the unique equilibrium. We thus follow the literature and adopt numerical techniques to solve the dynamic programming problem [See Prescott and Visscher (1977) as well as Lane (1980)].

\section{Equilibrium Location Choices}\label{sec:equilibrium}

In the location game, the equilibrium depends on the ratio between fixed costs, $F$, and market size, $M$. We choose to set $F=25$ and vary the market size. The results show firm locations shifting to deter entry as the market size increases. These shifts are not continuous nor are they symmetric. As has been noted since Eaton and Lipsey (1975), when given the opportunity, firms strategically use location (in physical or product space) to deter entry of competing firms.

In the interest of brevity, we limit our presentation to two characteristic location equilibria for each set of firms. The first is immediately after a firm enters and there is no threat of, and therefore no need to deter, a new entrant. The second is immediately prior to a new entrant, meaning there is a need to deter entry. The locations for 1, 2, and 3 firms in two dimensions are the same as they would be in one dimension. The locations become more interesting when we move to 4 and more firms.

\subsection{One, Two, and Three Active Firms}\label{sec:smallN}
\begin{figure}[h]
  \centering
     \caption{One Firm with Threat of Entry by Second Firm}    
	\includegraphics[width=0.33\textwidth]{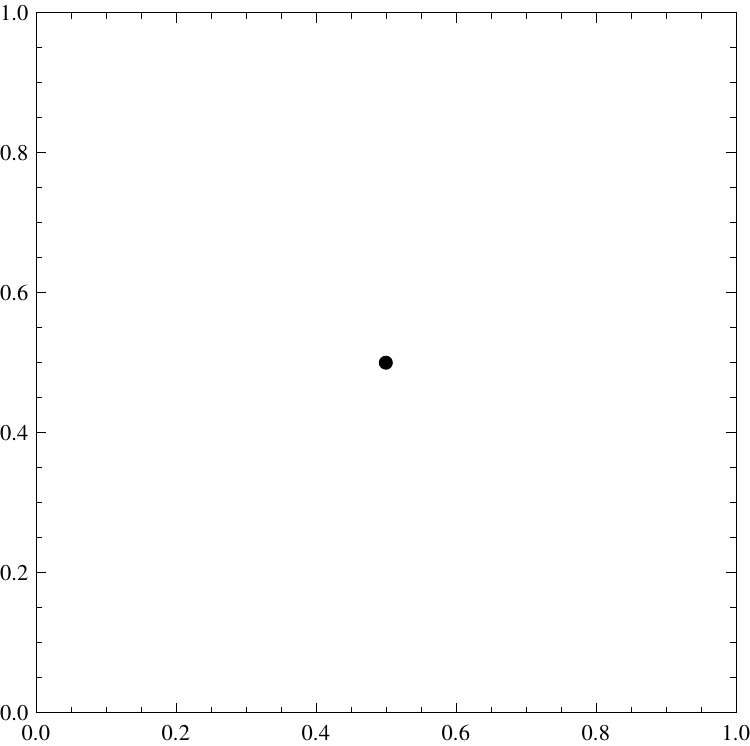}
   \label{fig1}
\end{figure}

Initially, a single firm can locate anywhere in a square market for small $M$. Eventually, as the density of the market increases (or equivalently, the size of the market grows), the firm will have to locate in the center of the market to deter entry of a second firm. $x_1 = (1/2, 1/2)$. See Figure \ref{fig1} at back.

\begin{figure}[h]
  \centering
     \caption{Two Firms with No Threat of Entry by Third Firm}    
	\includegraphics[width=0.33\textwidth]{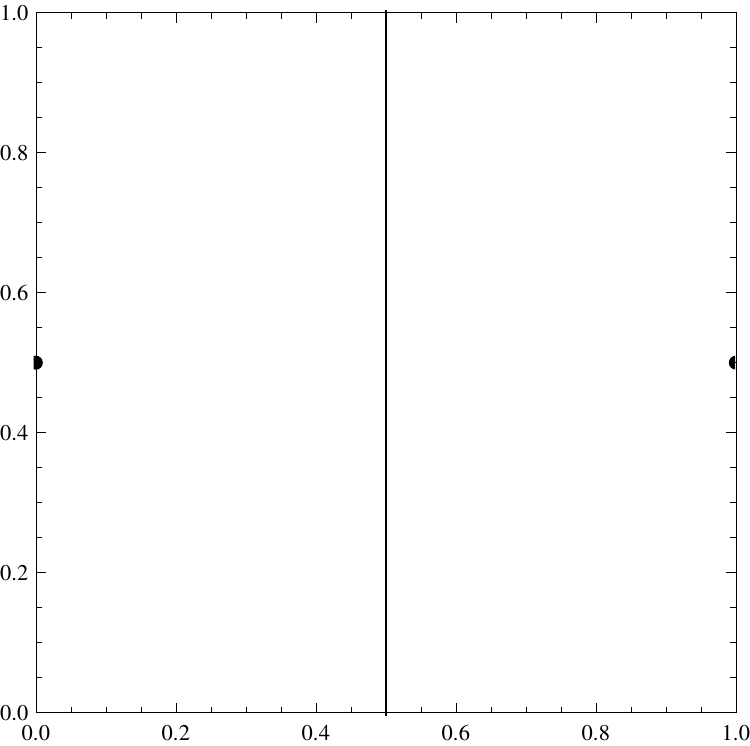}
   \label{fig2}
\end{figure}

\begin{figure}
  \centering
     \caption{Two Firms with Threat of Entry by Third Firm}    
	\includegraphics[width=0.33\textwidth]{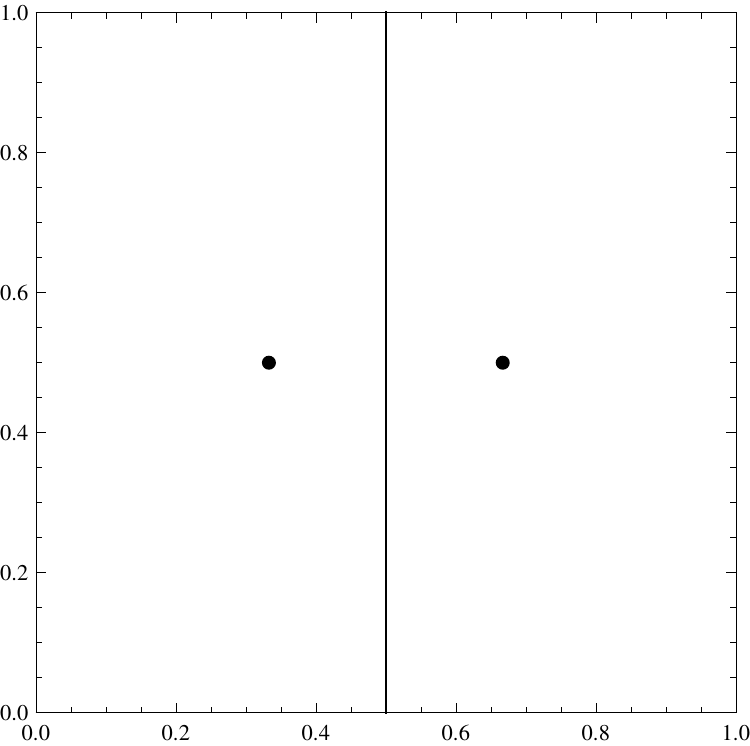}
   \label{fig23}
\end{figure}

While in the Hotelling model, firms locate next to each other in the center of the market, in two dimensions they locate as far away as possible so that they can minimize price competition. However, this is only the case when there is no threat of a third firm entering. The two firms locate closer and closer to each other as they balance the need to maximize market space with the need to deter entry of a third firm. Eventually the two firms locate at the thirds. These, then, are the threshold cases. Case 1 is when a second firm can just enter the market and firms locate at far ends, such as $x_1 = (0, 1/2)$ and $x_2 = (1, 1/2)$. Note that there is no unique location equalibrium but rather a set of possible location equilibrium. This set contains the borders of the market with the second firm locating the maximum distance from the first firm's location. [See Figure \ref{fig2} for an example]. Case 2 is immediately prior to entry by a third firm.  In this case, firms are located at the thirds on one axis and at the half on the other, $x_1 = (1/3, 1/2)$ and $x_2 = (2/3, 1/2)$. See Figure \ref{fig23} at back.

\begin{figure}
  \centering
     \caption{Three Firms with No Threat of Entry byFourth Firm}    
	\includegraphics[width=0.33\textwidth]{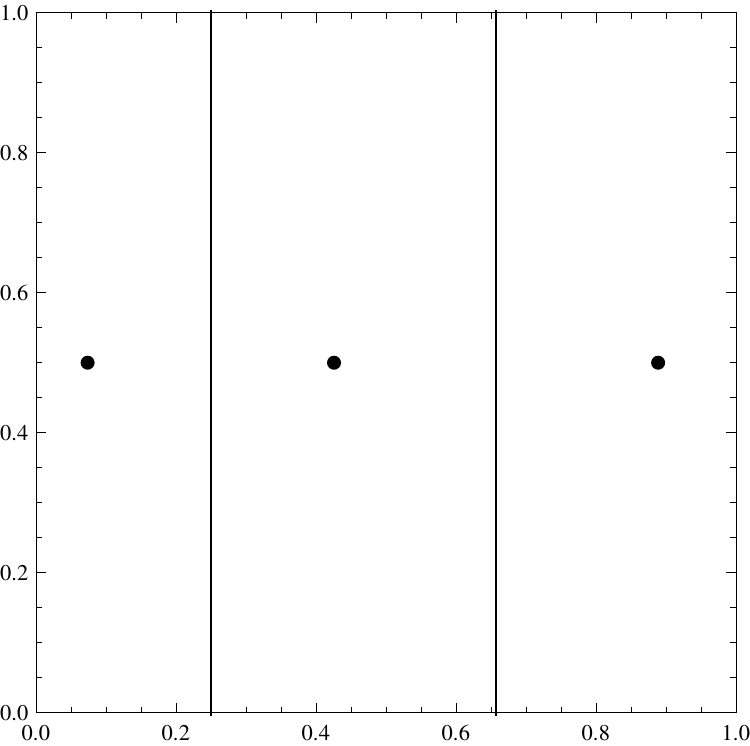}
   \label{fig3}
\end{figure}

When considering three active firms, we again have two threshold cases to consider. One when the third firm just enters the market. The other when a fourth firm is about to enter but has yet to enter. When a third firm first enters the market, the locations are $x_1 = (.426, .5), x_2 = (.889, .5), x_3 = (.074, .5)$. [See Figure \ref{fig3}]. While in the previous cases, firms located symmetrically and earned equal profits, once a third firm enters, rents accrue to early entrants. In this case, firm 1 earns excess profits due to the firm's first mover status. The firm is able to choose a location that forces the second entrant to locate at the far side of the market and induces the third entrant to locate in the its "backyard." Thus, firm 1 is able to secure a market share larger than the subsequent entrants.

\begin{figure}
  \centering
     \caption{Three Firms with Threat of Entry by Fourth Firm}    
	\includegraphics[width=0.33\textwidth]{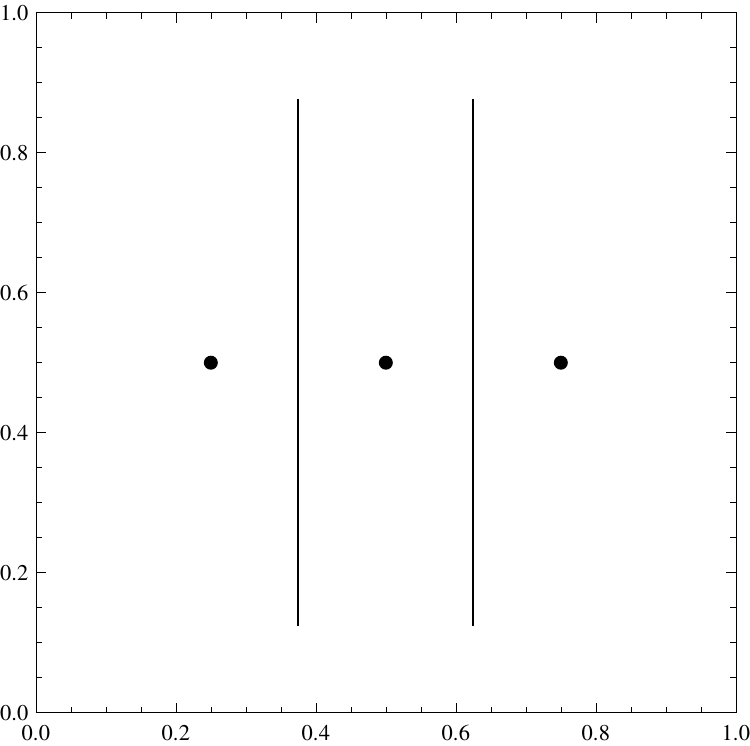}
   \label{fig34}
\end{figure}

As it becomes possible for a fourth firm to enter, all three firms attempt to defer entry by locating towards the quarter points. At the point where the three firms cannot stop a fourth firms from entering, they would be at the following locations: $x_1 = (1/4, 1/2), x_2 = (3/4, 1/2), x_3 = (1/2, 1/2)$. [See Figure \ref{fig34}]. The first firm is no longer in the central location but chooses an outside location as does the second firm. The third firm would prefer to choose a location as far away from firm 1 and 2 as possible (say $(1/2, 1)$) in order to maximize its market share and minimize price competition. However, the need to deter entry by a fourth firm results in firm 3 locating between firm 1 and 2. The benefit of such a location for firm 3 is that is blocks entry by firm 4, guaranteeing it higher profits than if it had located in the hinterland and allowed entry by a fourth firm.

\subsection{Four, Five, and Six Active Firms}\label{sec:largeN}
When a fourth firm enters the market, the equilibrium characteristics change. With 1-3 firms, firms strove for maximum differentiation on one axis why accepting minimum differentiation on the other axis. This is due less to strategic decision making on the part of the firms and more to the geometry of the space. The locations of two firms can always be connected by a single vector and thus can never achieve differentiation along both axes. A third firm could locate off of the vector connecting the first two firms yet these locations turn out not to be profit maximizing.

\begin{figure}
  \centering
     \caption{Four Firms with No Threat of Entry by Fifth Firm}    
	\includegraphics[width=0.33\textwidth]{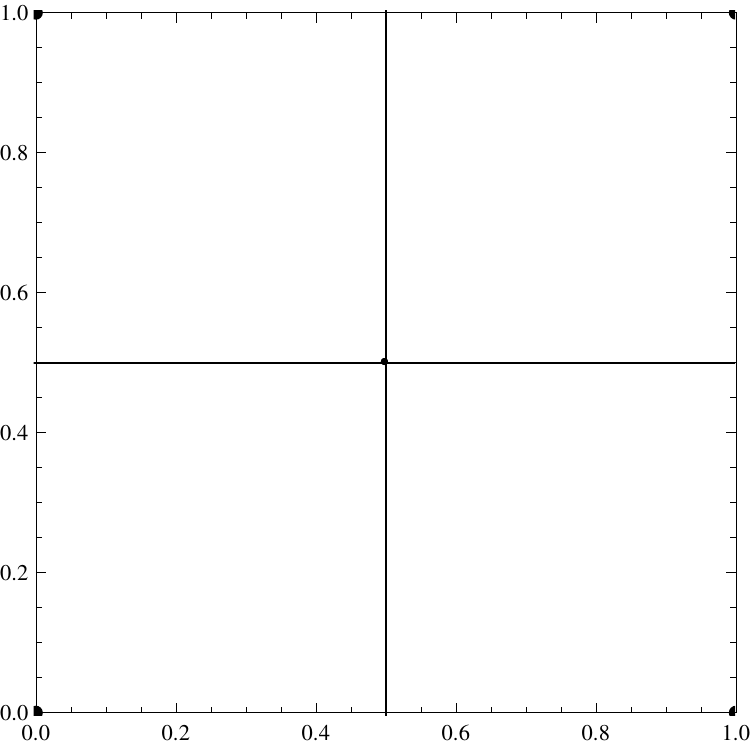}
   \label{fig4}
\end{figure}

\begin{figure}
  \centering
     \caption{Four Firms with Threat of Entry by Fifth Firm}    
	\includegraphics[width=0.33\textwidth]{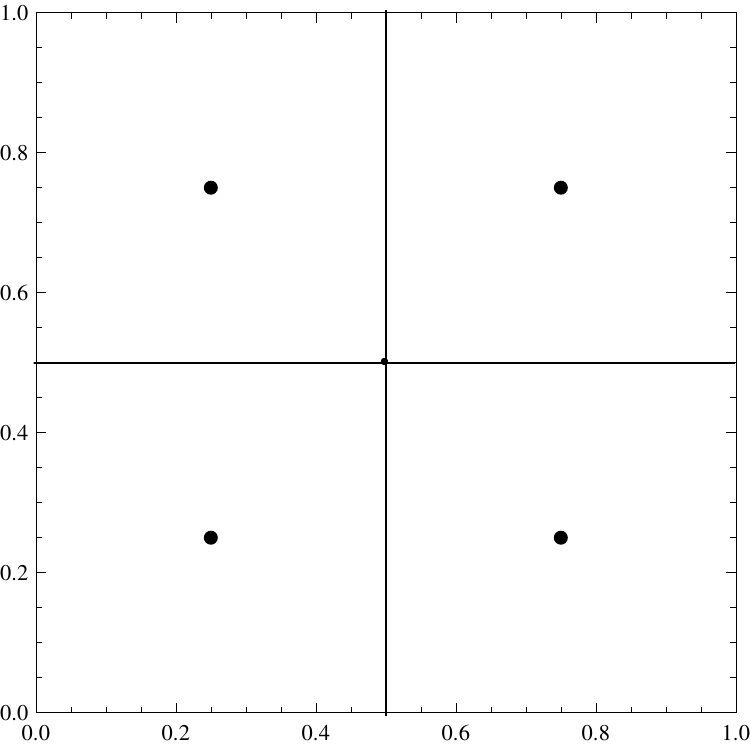}
   \label{fig45}
\end{figure}

With the immediate entry of a fourth firm, each firm locates at the corners of the market space: $x_1 = (0, 0), x_2 = (1, 1), x_3 = (0, 1), x_4 = (1,0)$. [See Figure \ref{fig4}]. This minimizes price competition by maximizing the distance between each firm. As the need arises to deter the entry of a fifth firm, the first four firms locate in positions that divide the market equally into quarters:  $x_1 = (1/4, 1/4), x_2 = (3/4, 3/4), x_3 = (1/4, 3/4), x_4 = (3/4,1/4)$. [See Figure \ref{fig45}]. As was the case with two firms, locations in the four firm case are symmetric, dividing the market space equally. This result is not only unique to even numbers of competing firms, it is unique to $N=2,4$. As will be seen, the symmetric location choices hold in $N=6$ yet these locations no longer divide the space equally. We conjecture that symmetric locations are a characteristic for all games with even numbers of firms competing while only equal profits for all firms only arise in $N=2,4$.

\begin{figure}
  \centering
     \caption{Five Firms with No Threat of Entry by Sixth Firm}    
	\includegraphics[width=0.33\textwidth]{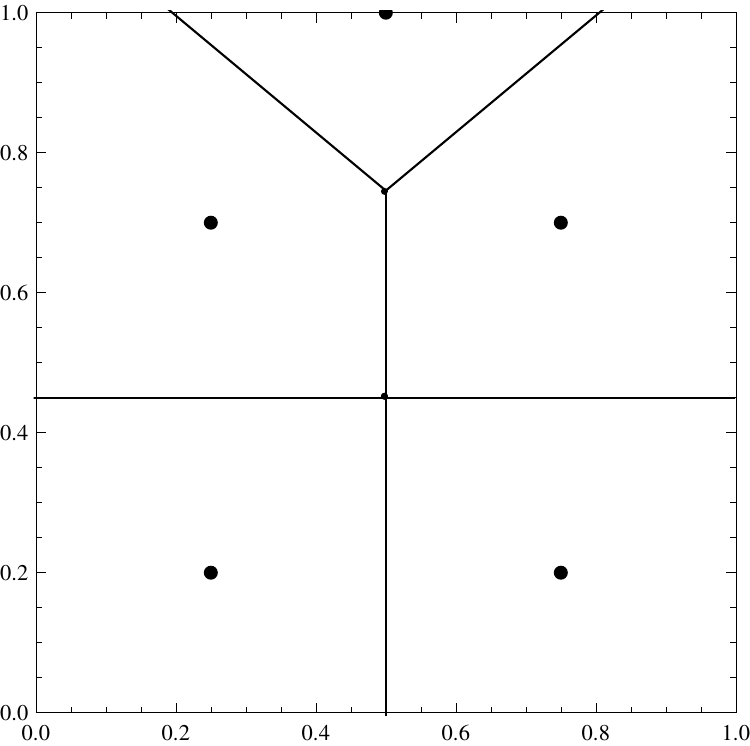}
   \label{fig5}
\end{figure}

\begin{figure}
  \centering
     \caption{Five Firms with Threat of Entry by Sixth Firm}    
	\includegraphics[width=0.33\textwidth]{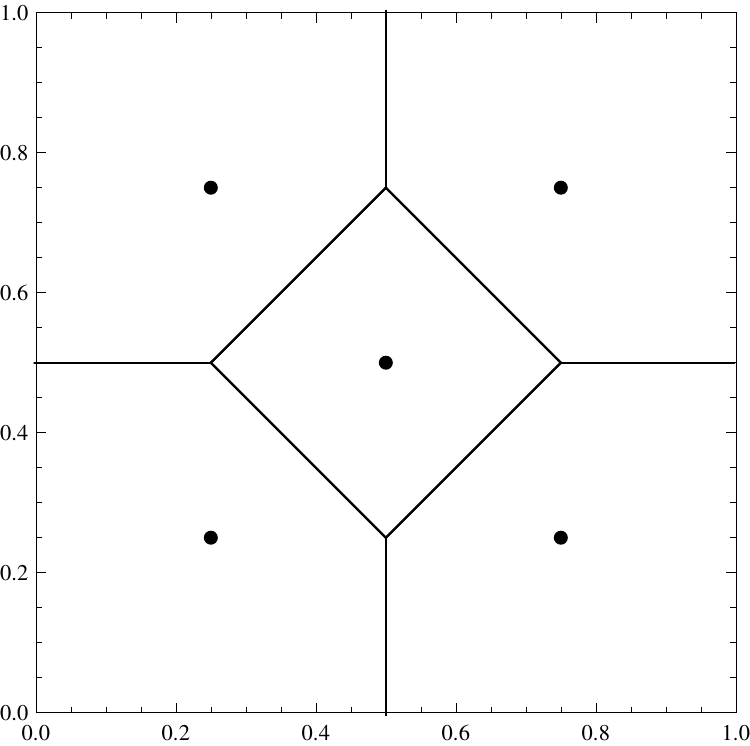}
   \label{fig56}
\end{figure}

When a fifth firm is just able to enter, the four firms would adopt slightly "asymmetric" locations, as occurred with firm 1 and 2 when firm 3 was just able to enter. In this case, firm 1 and 2 select more profitable locations by shading towards one market boundary. This ensures that firm five does not locate in their neighborhood but in the neighborhood of firm three and four: $x_1 = (.25, .15), x_2 = (.75, .15), x_3 = (.25, .65), x_4 = (.75,.65), x_5 = (.5,1)$. [See Figure \ref{fig5}]. When a sixth firm is about to enter, the five competing firms locate similar to the fifth face of a six-sided die. As with previous cases, firm accept stiffer price competition amongst themselves as the cost of deterring the entry of another firm. While the first four firms are able to secure equal market sizes, firm five, as the last entrant, receives less profit. Locations are $x_1 = (1/4, 1/4), x_2 = (3/4, 3/4), x_3 = (1/4, 3/4), x_4 = (3/4,1/4), x_5=(1/2,1/2)$. [See Figure \ref{fig56}.

\begin{figure}
  \centering
     \caption{Six Firms with No Threat of Entry by Seventh Firm}    
	\includegraphics[width=0.33\textwidth]{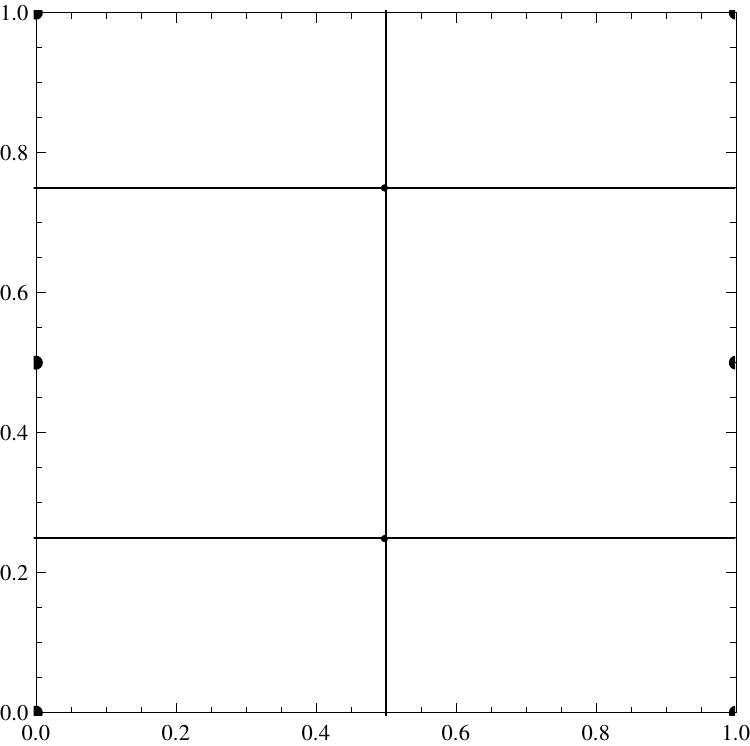}
   \label{fig6}
\end{figure}

\begin{figure}
  \centering
     \caption{Six Firms with Threat of Entry by Seventh Firm}    
	\includegraphics[width=0.33\textwidth]{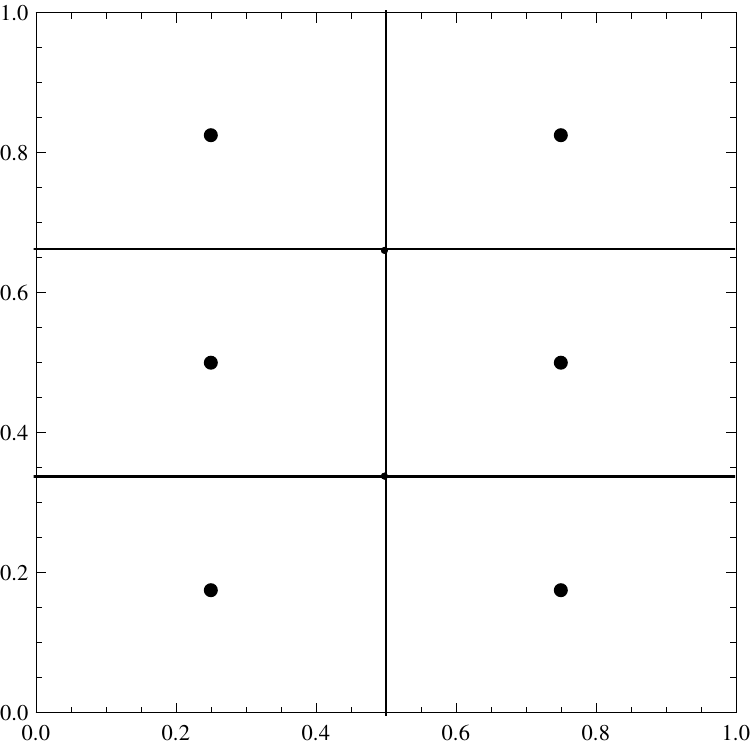}
   \label{fig67}
\end{figure}

When a sixth firm just enters the market, all firms locate along the borders to try and maximize their distance from each other: $x_1 = (0, 1/2), x_2 = (1, 1/2), x_3 = (0,1), x_4 = (1,0), x_5 = (0,0), x_6 = (1, 1)$. [See Figure \ref{fig6}]. As mentioned above, though the locations are symmetric, the resulting market size is asymmetric. Firm 1 and 2 locate as they did in the $N=2$ case, forcing the subsequent four firms to take locations in the corners of the market. The first two entrants earn rents from their early entry into the market. Subsequent firms are indifferent regarding which corner they locate in, since they cannot earn more profits in one corner compared to another. As it became necessary to deter entry of firm 7, the first 6 firms locate in the interior of the market [See Figure \ref{fig67}].

A curious result is that firms 3,4,5 and 6 prefer this configuration to the initial configuration when there was no chance of firm 7 entering. A similar asymmetry in the burden of deterrence can be seen to a lesser extent in the $N=5$ firm case. This asymmetry in who bears the burden of deterrence was first noticed in the one dimensional case by Gotz (2005). 

\section{Comparison of Results} \label{sec:results}
Several new insights emerge when we examine endogenous sequential entry in two dimensions instead of just one. First, for small $N$, specifically $N=1,2,3$, optimal locations in two dimensions are equivalent to optimal locations in one dimension. Numerous papers that examine endogenous sequential entry in one dimension, most prominently Lane (1980) and Veendorp and Majeed (1995), have conjectured that little insight is to be gained by expansion of the location problem into two dimensions. Our results show that this conjecture, while accurate for $N=1,2,3$ is erroneous for $N \geq 4$. Something similar can be found in Okabe and Suzuki (1987) for the two dimensional simultaneous entry problem. Eaton and Lipsey (1975) had conjectured that no stable equilibria exist when firms enter simultaneously into a two dimensional market. Shaked (1975) had proven this for up to three firms. It was not till Okabe and Suzuki (1987) that it was demonstrated stable equilibria do exist for large $N$. In our problem, the equivalence between locations in one and two dimensions for 1, 2, or 3 firms does not carry over for more than 3 firms. We have examined the cases up to $N=7$ and observe that a pattern develops. However, we can only conjecture that this pattern carries beyond $N=7$.

Our second insight involves the asymmetry in the burden of entry deterrence. As noted above, when a firm has just entered the market, the earlier entrants are able to select preferred locations, earning rents, and forcing the new entrant into a less preferred location. These rents are diminished by the need to deter entry by subsequent firms. In their attempt to restrict competition to a small number of firms, more profitable firms sacrifice profit to less profitable firms, thereby increasing equity between the operating firms. Thus, late entrants prefer a market environment where the threat of additional entrants exist to one that is secure for further competition.

Our third insight involves the social optimality of the location equilibria. We define social optimality as the minimum total transportation cost:

$$F(\bf{X^*})=\min_{\bf{X}}F(\bf{X}) \eqno (6)$$

\noindent where  $\bf{X^*}=(\bf{x_1^{*^T}},\bf{x_2^{*^T}},..., \bf{x_n^{*^T}})^T$. The problem with such a concept is that $F(\bf{X^*})$ is nonconvex so there is not a unique optimal solution. This is because from the consumer's perspective $(\bf{x_1^{*^T}},\bf{x_2^{*^T}},...,\bf{x_n^{*^T}})^T=(\bf{x_2^{*^T}},\bf{x_1^{*^T}},..., \bf{x_n^{*^T}})^T$. Since all firms are homogenous, consumers are indifferent between which firm they buy from. Not only are socially optimal equilibria not unique, they are generally not the type of equilibria that arise when firms solve their profit maximization problem. The only equilibria that are socially optimal are those that arise to deter entry of an additional firm. Thus, the threat of a new entrant results in both a more equitable distribution of profits and a socially optimal configuration of firms.

Finally, the way that firms pursue differentiation in the market is fundamentally different in two dimensions than it is in one dimension. If we think of product characteristic space instead of physical space, these results have very intuitive interpretations. When firms in a market do not have to worry about the entry of an additional firm they pursue maximum product differentiation. But, when there is the potential for a new firm to enter the market, the firms "move" or "locate" their products closer to each other. They do no try to pursue minimum product differentiation but rather accept greater similarity in product characteristics to ward off further entry. Put another way, when there is no threat of a new product coming into the market, existing firms trumpet how different their product is from those of the other operating firms. When the threat of a new product entering the market arises, firms shift their strategy to try and "fill in" the product space. Where before they "advertised" how their product was different from all others, now they "advertise" how their product has similar characteristics to all the others. Basically, no firm wants to leave a gap in the market where a new firm could enter so all existing firms produce products with some of each characteristic. This ensures that a new firm cannot enter into the market by claiming "buy our product because it has a full compliment of $x$ while that other product has zero for $x$." Rather, to keep that new entrant out, existing firms produce a product that has some level of $x$ so a new entrant cannot slip into an unexploited product characteristic zone.

\section{Conclusion} \label{sec:conclusion}
Little beyond intelligent conjecture has been done on location models in two dimensional space, especially when entry is sequential and the number of firms is endogenous to the market. This gap in the literature is curious for two reasons. First, conjectures regarding equilibria in two dimensions for simultaneous entry were proven to be incorrect (Okabe and Suzuki, 1987). This paper has attempted to examine conjectures regarding equilibria in two dimensions when entry is sequential. While the conjectures put forth by the literature are accurate for $N=1,2,3$ firms, they are inaccurate for larger numbers of firms. Second, most products have characteristics along more than one dimension and a model of how firms choose location in two dimensions could prove useful. Our results show that the two dimensional model leads to insights not present in the one dimensional model.

Among the insights that arise from our model we believe the most interesting are the asymmetric burden of deterrence and the shifting approach to product differentiation. The first insight provides two dimensional verification for a phenomenon first observed by Gotz (2005) in a single dimension. If we think of the growth of the market space ($N$) as a dynamic process, then early entrants capture rents in the form of higher profits. Yet, these rents are eroded as the market grows and the need to deter new entrants arises. There is a move towards a more equitable outcome among active firms in an attempt to prevent more intense competition. The analogy is not completely accurate yet is suggestive. The second insight fails to verify the results presented in Veendorp and Majeed (1995). In their model of a two dimensional market they found firms minimized differentiation along one dimension while maximized differentiation along a second dimension. While this is the case when the market has 1, 2, or 3 active firms, it is not when $N \geq 4$. We find that absent the threat of a new entrant, firms attempt to maximize differentiation along all dimensions. Only when a new entrant is imminent do we find firms attempt to "fill in" the gap in the market space by reducing differentiation along one dimension.

The model does not lead, generally, to socially optimal equilibria. Even as the number of firms in the market increase, thus increasing competition, a socially optimal outcome is not guaranteed. Obviously, for extremely large $N$ there will be so many firms in the market that the difference between the achieved equilibrium and the socially optimal equilibrium will be trivial. Yet, it is worth repeating that in a two dimensional market where firms choose location and price, increased competition does not necessarily lead to socially optimal equilibria. The fixed cost of entry, the element in the models of Hay (1976) and Rothschild (1976) that give rise to socially optimal outcomes on the Hotelling Line, appears to allow firms to capture rents from early entry into the market.

\newpage
\section*{References}
\addcontentsline{toc}{section}{References}
\singlespacing

\noindent Economides, N., Howell, J., and S. Meza. 2002. "Does it Pay to be First? Sequential

Location Choice and Foreclosure." Stern School of Business, New York University.

EC-02-19.

\noindent Eiselt, H.A. and G. Laporte. 1996. "Sequential Location Problems." \emph{European Journal of}

\emph{Operational Research} 96: 217-31.

\noindent Eiselt, H.A., Laporte, G., and J.-F. Thisse. 1993. "Competitive Location Models: A

Framework and Bibliography." \emph{Transportation Science} 27: 44-54.

\noindent Gotz, G. 2005. "Endogenous Sequential Entry in a Spatial Model Revisited." \emph{International}

\emph{Journal of Industrial Organization} 23: 249-61.

\noindent Hakimi, S.L. 1983. "On Locating New Facilities in a Competitive Environment." \emph{European}

\emph{Journal of Operational Research} 12: 29-35.

\noindent Hay, D.A. 1976. "Sequential Entry and Entry-Deterring Strategies in Spatial Competition." 

\emph{Oxford Economic Papers} 28: 240-57.

\noindent Hotelling, H. 1929. "Stability in Competition." \emph{Economic Journal} 39: 41-57.

\noindent Lane, W.J. 1980. "Product Differentiation in a Market with Endogenous Sequential Entry."

\emph{Bell Journal of Economics} 11: 237-60.

\noindent Neven, D.J. 1987. "Endogenous Sequential Entry in a Spatial Model." \emph{International Journal}

\emph{of Industrial Organization} 5: 419-34.

\noindent Okabe, A. and A. Suzuki. 1987. "Stability of Spatial Competition for a Large Number of 

Firms on a Bounded Two-Dimensional Space." \emph{Environment and Planning A} 19: 

1067-1082

\noindent Prescott, E.C. and M. Visscher. 1977. "Sequential Locatio Among Firms with Foresight." 

\emph{Bell Journal of Economics} 8: 378-93.

\noindent Rothschild, R. 1976. "A Note on the Effect of Sequential Entry on Choice of Location." 

\emph{Journal of Industrial Economics} 24: 313-20.

\noindent Shaked, A. 1975. "Non-Existence of Equilibrium for the Two-Dimensional Three-Firms

Location Problem." \emph{Review of Economic Studies} 42: 51-5.

\noindent Veendorp, E.C.H. and A. Majeed. 1995. "Differentiation in a Two-Dimensional Market." 

\emph{Regional Science and Urban Economics} 25: 75-83.

\mdseries

\newpage
\section*{Figures}
\addcontentsline{toc}{section}{Figures}

\end{document}